\documentclass[apjl]{myemulateapj}
\usepackage{apjfonts}

\shorttitle{Polarimetric observations of $\sigma$ Orionis E}
\shortauthors{Carciofi et al.}

\begin{document}
\title{Polarimetric observations of $\sigma$ Orionis E}

\author{A. C. Carciofi, D. M. Faes}
\affil{Instituto de Astronomia, Geof\'isica e Ci\^encias Atmosf\'ericas, Universidade de S\~ao Paulo, Rua do Mat\~ao 1226, Cidade Universit\'aria, 05508-900, S\~ao Paulo, SP, BRAZIL}
\email{carciofi@usp.br}
\author{R. H. D. Townsend}
\affil{Department of Astronomy, University of Wisconsin-Madison, Sterling Hall, 475 N. Charter Street, Madison, WI 53706, USA}
\and
\author{J. E. Bjorkman}
\affil{Ritter Observatory, Department of Physics \& Astronomy, University of Toledo,Toledo, OH 43606, USA}

\begin{abstract} 
Some massive stars possess strong magnetic fields that confine plasma in the
circumstellar environment. These \textit{magnetospheres} have been studied
spectroscopically, photometrically and, more recently, interferometrically.
%PURPOSE+%METHOD
Here we report on the first firm detection of a magnetosphere in continuum linear polarization, as a result of monitoring of $\sigma$\,Ori\,E at the Pico dos Dias Observatory. {The non-zero intrinsic polarization indicates an asymmetric structure, whose minor elongation axis is oriented $150\fdg0$ east of the celestial north.}
A modulation of the polarization was observed, with a period of half of the rotation period, which supports the theoretical prediction of the presence of two diametrally opposed, co-rotating blobs of gas.
A phase lag of $-0.085$ was detected between the polarization minimum and the primary minimum of the light curve, suggestive of a complex shape of the plasma clouds. %From the polarization angle, we were able to determine that the rotation axis of the star is oriented $150\fdg0$ east of the celestial north.
We present a preliminary analysis of the data with the Rigidly Rotating Magnetosphere model, which could not reproduce simultaneously the photometric and polarimetric data. A toy model comprising two spherical co-rotating blobs {joined by a thin disk} proved more successful in reproducing the polarization modulation. 
{With this model we were able to determine that the total scattering mass of the thin disk is similar to the mass of the blobs ($2M_{\rm b}/M_{\rm d}=1.2$) and that the blobs are rotating counterclockwise on the plane of the sky.}
This result shows that polarimetry can provide a diagnostic of the geometry of clouds, which will serve as an important constraint for {improving} the Rigidly Rotating Magnetosphere model.
\end{abstract}


%Cost of Printed color page (per printed page) = $350, since April 2011
%http://aas.org/journals/authors/publication_charges

%IMPORTANT:
%WHat are the effects of the companion on the polarization? See Rich's and Vero's paper in CSDYN

\section{Introduction}
 \label{intro_aca}

%It has been recently found that several early-type stars exhibits strong magnetic fields, whose origin is not yet known. In some cases, the field is so intense that it can trap (part of) the stellar wind outflowing material, thereby producing a magnetosphere that co-rotates with the star \citep{1997A&A...323..121B}.

{The helium-strong star $\sigma$\,Orionis\,E (HD\,37479; B2\,Vpe; $m_V=6.66$) has long been known to possess a circumstellar magnetosphere that is formed by wind plasma that is  trapped by a strong dipolar magnetic field \citep[$\approx$10 kG, ][]{1978ApJ...224L...5L}. 
The presence of this magnetosphere can be inferred by the eclipse-like dimmings on its light curve, which is thought to occur when plasma clouds go in front of the star twice every rotation cycle \citep{2005ApJ...630L..81T,1977ApJ...216L..31H,1982A&A...116...64G}.}
This star has been the testbed for important advancements in our theoretical understanding of these magnetospheres. The Rigidly Rotating Magnetosphere (RRM) models of this star were successful in reproducing the observed variability in emission line profiles and photometry \citep{2005ApJ...630L..81T}.

Polarization is a very useful technique that allows one to probe the geometry of the circumstellar scattering material without angularly resolving it \cite[e.g.,][]{1977A&A....57..141B}. \citet{1977ApJ...218..770K} carried out the first polarimetric observations of the $\sigma$\,Ori system, but their results where largely inconclusive because of the small values of the intrinsic polarization.
In this paper we present the results of high-precision ($\sigma \sim 0.01$\%) polarization monitoring of $\sigma$\,Ori\,E, that has resulted in the first firm detection of the polarization modulation produced by a co-rotating magnetosphere.

\section{Observations}
%polarimetric data description
Broad-band linear polarization data were obtained  from August, 2010 to September, 2011, using the IAGPOL image polarimeter attached to the 0.6-m Boller \& Chivens telescope at Pico dos Dias Observatory (OPD/LNA).  %A summary of our polarimetric data is presented in Table \ref{tab:vdata}.
We used a CCD camera with a polarimetric module described in \citet{1996ASPC...97..118M}, consisting of a rotating half-waveplate and a calcite prism placed in the telescope beam. 
In each observing run at least one polarized standard star was observed in order to calibrate the observed position angle. HD\,23512 and HD\,187929 were used as polarized standards.

Although typical polarimetric observations consists of 8 consecutive wave plate positions (hereafter WPP) separated by 22\fdg5 \citep{2007ApJ...671L..49C}, in this work we chose to use 16 consecutive WPPs in order to increase the signal-to-noise ratio. 
Also, to reach the high accuracy needed to monitor the modulation of the polarization, a large number of frames (typically from 20 to 100) were obtained at each WPP.
Since the  exposure times ranged from 0.3 to 1\,s per frame, the  temporal resolution of our observations vary from 7 to 34\,min, including telescope and instrument overheads.  
These values are short enough to temporally resolve the polarimetric variation over the 1.19\,d period of the star.
Details of the data reduction can be found in \citet{1984PASP...96..383M}.

\begin{figure*}[!t]
\centering
\includegraphics[width=.27\linewidth,clip=true]{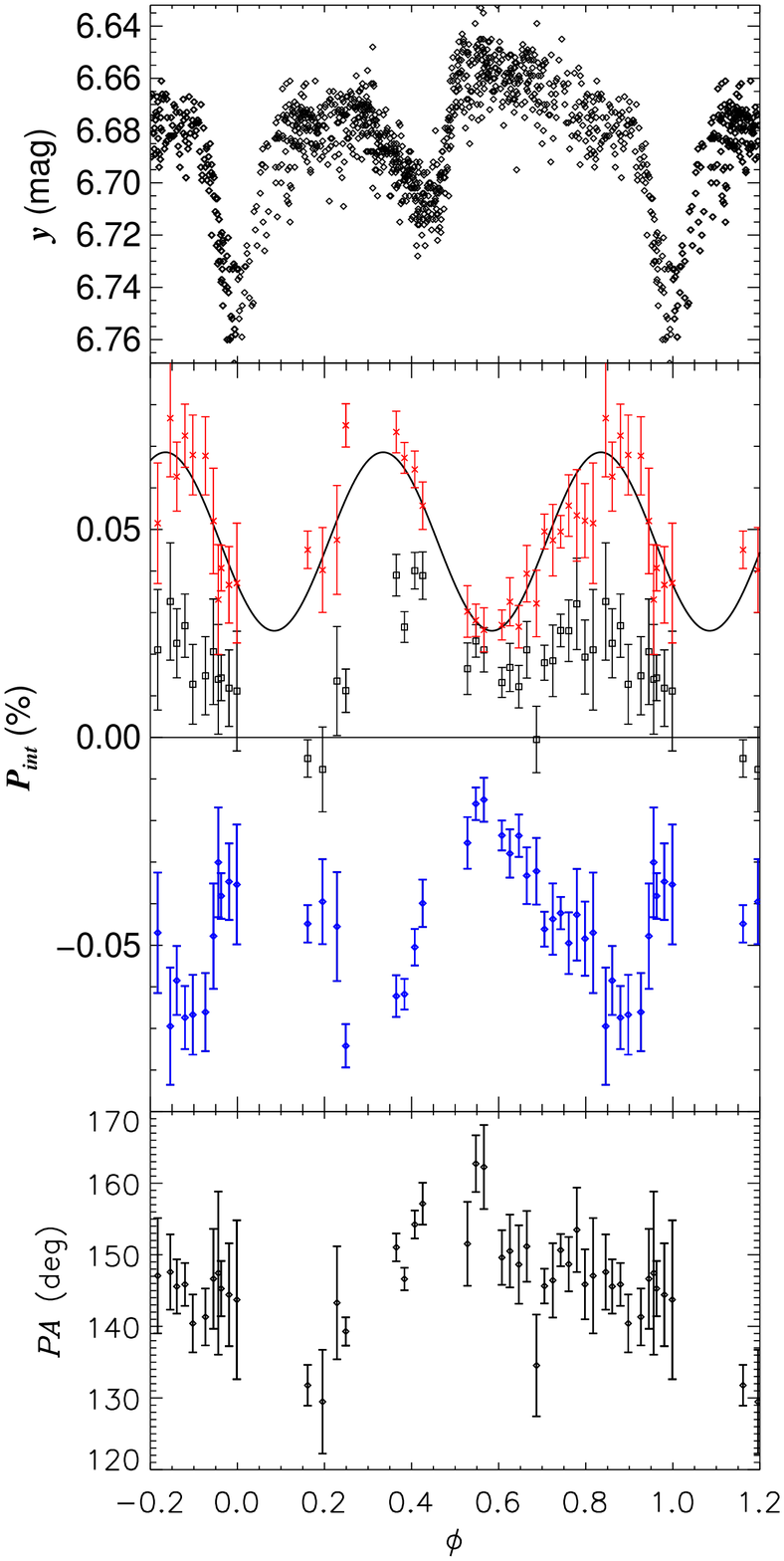}
\includegraphics[width=.27\linewidth,clip=true]{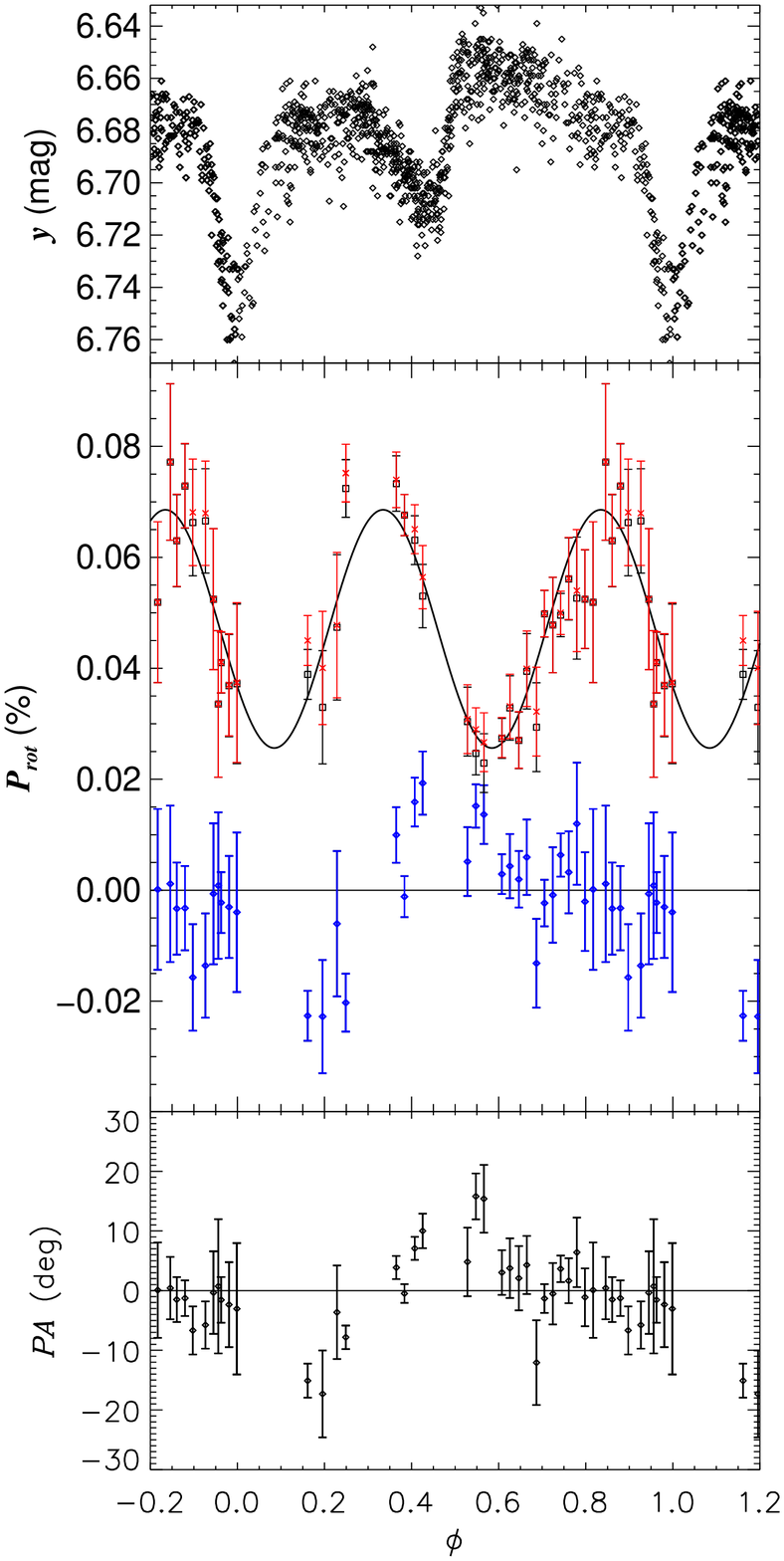}
\includegraphics[width=.27\linewidth,clip=true]{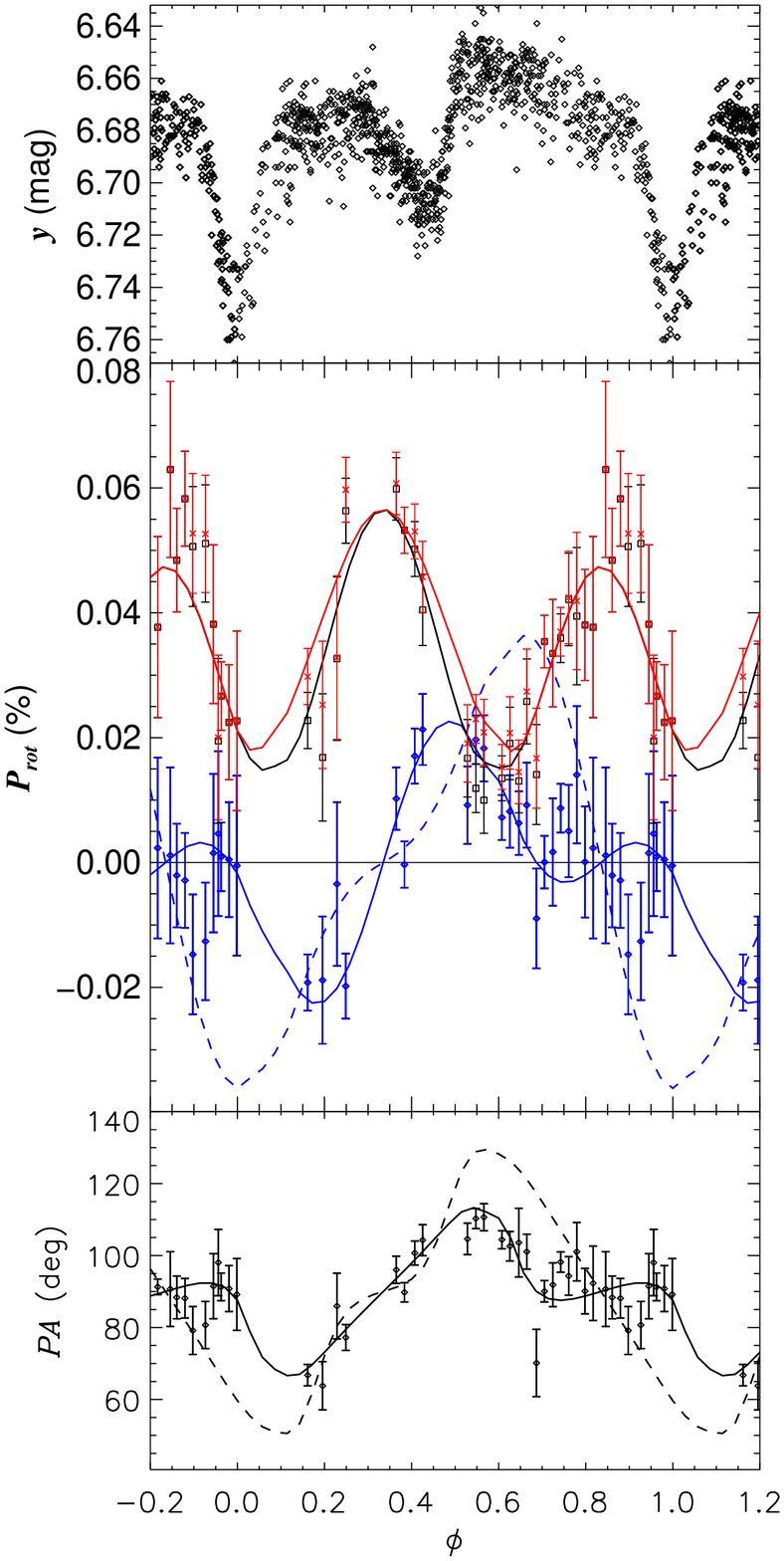}
\caption{Variation of the intrinsic polarization of $\sigma$\,Ori\,E compared with photometric variations in the $y$ filter. 
\textit{Top panels:} $y$ filter photometric data from \citet{hes77}, folded in phase. \textit{Middle panels:} intrinsic $P$ (red), $Q$ (black) and $U$ (blue). \textit{Bottom panels:} intrinsic position angle. 
{\it Left plot:} The polarization data is displayed in the equatorial reference frame. The polarization of $\sigma$\,Ori AB was used as a measure of the IS polarization.
{\it Middle plot: } Same as the left plot, rotated such that $\langle U_{\rm int} \rangle = 0$.
{\it Right plot: } Polarization compared to the single-scattering %``dumbbell'' 
model.  The solid line is for our best-fitting model with $i=70^\circ$ and the dashed line is for $i=110^\circ$. 
For this plot we used the estimate for the IS polarization and $\theta$ made in Sect.~\ref{sec:dumbbell}.
 \label{fig:intrinsic}}
\end{figure*}

\section{Interstellar Polarization}

\citet{1977ApJ...218..770K} report similar values of the polarization for $\sigma$\,Ori\,AB, $\sigma$\,Ori C and $\sigma$\,Ori D, which suggests that all these stars  belong to the same physical system, and that the measured polarization is of interstellar (IS) origin.

We measured the $BVRI$ polarization of $\sigma$\,Ori\,AB and C and our results are largely consistent with the ones of  \citet{1977ApJ...218..770K}. 
%In Table~\ref{tab:ISdata} we show the average polarization of $\sigma$\,Ori\,AB from the observations made between August 17th and 19th, 2010.  These results were used as a measure of the foreground polarization towards $\sigma$\,Ori\,E.
We used our measured $V$-band polarization of $\sigma$\,Ori\,AB ($P=0.351(15)$\%; 86.7$^{\circ}$; $Q=-.348(15)$\%; $U=0.040(15)$\%) as an estimate of the IS polarization towards $\sigma$\,Ori\,E.

\section{Results} \label{results_aca}

%The $V$-band $\sigma$\,Ori\,E observations %described in Table~\ref{tab:vdata} 
%are shown in Fig.~\ref{fig:observed}. \addACC{aqui} The data was folded in phase ($\phi$) using the ephemerides of \citet{2010ApJ...714L.318T}. The IS polarization is also shown in the plot as the dashed horizontal lines. Whilst the $\sigma$\,Ori\,E polarization is mostly of IS origin, as seen by the fact that both $Q_{\rm obs}$ and $U_{\rm obs}$ are similar to the IS values, there is a clear modulation of the data, indicative of an intrinsic polarization of $\sigma$\,Ori\,E.

%%%%%%%%%%%%%%%%%%%%%%%%%%
%\begin{figure}[!t]
%%\centerline{
%%\plotone{../figure1.eps}%}
%\centering
%\includegraphics[width=.70\linewidth]{f1}
%\caption{Observed polarization of $\sigma$\,Ori\,E. \textit{Top panel:} observed $P$ (red), $Q$ (black) and $U$ (green). 
%\textit{Bottom panel:} observed position angle. The data was folded on phase using the ephemeris of \citet{2010ApJ...714L.318T}.
%  \label{fig:observed} }
% %\end{center}
% \end{figure}

The $V$-band $\sigma$\,Ori\,E observations are shown in the left plot of Fig.~\ref{fig:intrinsic}. The data was folded in phase ($\phi$) using the ephemerides of \citet{2010ApJ...714L.318T}. Shown is the intrinsic polarization, calculated by subtracting the IS polarization, estimated using $\sigma$\,Ori\,AB, from the observed $Q$ and $U$ Stokes parameters.
%To bring out the intrinsic value of the polarization and its modulation, we subtracted the IS values from $Q_{\rm obs}$ and $U_{\rm obs}$, %(Table~\ref{tab:ISdata}), thus obtaining $Q_{\rm int}$ and $U_{\rm int}$ (Fig.~\ref{fig:intrinsic}, left). 
The polarization has a double-peaked structure, with maxima of about 0.07\% occurring at phases around 0.3 and 0.8 and minima occurring around phases 0.1 and 0.6. We note that the  values reported here are in broad agreement with the data of \citet{1977ApJ...218..770K}, but a more quantitative comparison is hampered by the insufficient accuracy of the latter data. 

There is a clear anti-correlation between the photometric and polarimetric variations, with the polarization minima roughly coinciding with the photometric maxima and vice-versa, but there are important differences between the two curves. One such difference lies in their symmetry: the photometry is clearly asymmetric, with the secondary minimum  occurring at about phase 0.4 (the primary minimum occurs at $\phi=0$ by definition), and the behavior of the two inter-minima  being very different from each other. The polarization curve, however, seems to be roughly symmetric with respect to $\phi\approx0.6$.
That the asymmetry is seen in the photometry, but not in the polarization, is further supporting evidence that -- as hypothesized in \citet{2005ApJ...630L..81T} -- the asymmetric light curve arises due to contamination of the circumstellar signal by a photospheric signal. That is, the photometric variations come from a combination of magnetospheric eclipses and photospheric abundance inhomogeneities, and therefore lack rotational symmetry; but conversely, because the polarization variations come solely from magnetospheric electron scattering, we do see rotational symmetry in their case.

In order to determine the location of the polarization minimum, we fitted the data between $\phi = 0.45$ and 0.75 with a parabola, using the Levenberg-Marquardt algorithm \citep{mar63}. 
We find that the minimum of the intrinsic polarization occurs at $\phi_{\rm min}=0.61\pm 0.03$. To determine the location of the maximum, the same procedure was applied to the data between $\phi=0.7$ and 1. According to this fit the maximum of the polarization occurs at $\phi_{\rm max}=0.84 \pm 0.04$. 
%The $\approx 0.23$ phase difference between maximum and minimum is suggestive of a period of one half of the orbital period.

To further assess the symmetry of the data we fitted the polarization data with the following function
\begin{equation}
P(\phi) = P_0 + A \cos[4\pi(\phi-\delta)]\,,
\label{eq:sin}
\end{equation}
which implicitly assumes a period of half of the orbital period for the polarimetric data.
The result of the fit gives $P_0 = 0.0471 \pm 0.0009\,\%$, $A = 0.021\pm 0.001\,\%$ and $\delta = -0.17\pm0.01$, with a reduced $\chi^2=0.67$. The fitted function is displayed as the black line in the middle plot of Fig.~\ref{fig:intrinsic}.

The data seems to be well represented by Eq.~(\ref{eq:sin}), which indicates
that, contrary to the photometry, the polarization curve is roughly
symmetric. From Eq.~(\ref{eq:sin}), the position of the minima are $\phi=0.08$
and 0.58, and the maxima are at $\phi=0.33$ and 0.83. Those values are in good
agreement with the phases derived above via a parabola fitting.  The minimum
and maximum values of the polarization are, respectively,
$P_0-A=0.026\pm0.001\%$ and $P_0+A=0.068\pm0.001\%$.

An important feature of the intrinsic polarization of $\sigma$\,Ori\,E is that it is never zero. This indicates that there is some degree of asymmetry of the scattering material (as seen projected onto the plane of the sky) throughout the entire rotation period.  

The data shown in the left plot of Fig.~\ref{fig:intrinsic} {are} in the equatorial frame. The weighted-average of all the  data gives a value of $\langle Q_{\rm int} \rangle =0.023\%$ and $\langle U_{\rm int} \rangle = -0.040\%$, which, in turn, results in an average position angle of $\langle \theta_{\rm int} \rangle =  150\fdg0$.  
This value tells us the (average) direction of the minor axis of the asymmetric structure. Furthermore, the small variation of the position angle (Fig.~\ref{fig:intrinsic}, left) tells us that the direction of the minimum elongation axis changes little ($\approx 10$--$15^\circ$, at most) as the star rotates.

This last point is better understood if we rotate the intrinsic polarization by  $\langle \theta_{\rm int} \rangle$ so that the average value of $U$ is zero (Fig.~\ref{fig:intrinsic}, right plot). In this new frame, we see that the amplitude of the rotational modulation of  $U$ is small. %, of the order of the typical error in our observations. Therefore, in order to have more precise information about the modulation of $U$ (and hence of the position angle), more precise data is necessary.  
If we assume that the symmetry axis is parallel to the rotation axis, %(Sect.~\ref{sec:modeling}), 
we conclude that the projected axis of symmetry of the magnetosphere varies little as the star rotates. As discussed below, this has quite important consequences for the RRM model.

Another noteworthy feature of the polarization curve is that there is a phase shift of $-0.085$ between the primary minimum of the lightcurve and the polarization minimum. The RRM model of \citet{2005ApJ...630L..81T} predicts the existence of two plasma clouds co-rotating with the star. If the clouds were symmetrical and the photospheric flux were homogeneous, there should be no phase difference between the polarization and photometric curves; the observed phase shift, therefore, is likely a result of cloud asymmetries and photospheric abundance inhomogeneities.

%The INTRINSIC polarimetric spectral dependence of Sigma Ori E => unfortunately the data is not so great.

%\begin{figure}[!t]
%%\centerline{
%%\plotone{../figure3.eps}%}
%\centering
%\includegraphics[width=.5\linewidth]{figure3.eps}
%\caption{Same as Fig.~\ref{fig:intrinsic}, but with the intrinsic polarization rotated such that  $\langle U_{\rm int} = 0 \rangle$. \label{fig:intrinsic_rotated} }
%\end{figure}

\begin{figure}[!t]
\centering
\includegraphics[width=.9\linewidth]{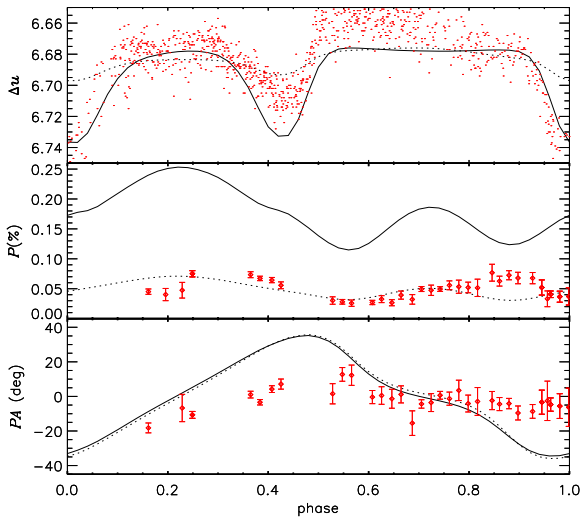}
\caption{Modeling of the intrinsic polarization of $\sigma$\,Ori\,E using the RRM model (observations are in red). The only free parameter is the maximum number density in the magnetosphere, which was set to $10^{12}\rm\, cm^{-3}$ (solid lines) to reproduce the depth of the eclipses and $2.5\times 10^{11}\rm\, cm^{-3}$ (dotted lines) to reproduce the amplitude of the linear polarization.
%The solid line corresponds to a viewing angle of $70^{\circ}$ and the dotted line to $110^{\circ}$.
\label{fig:rrm_fit} }
\end{figure}

\section{RRM Model \label{sec:modeling}}
\citet{2005ApJ...630L..81T} applied the RRM model to $\sigma$\,Ori\,E with good success. This model predicts the accumulation of circumstellar plasma in two co-rotating clouds, situated in magnetohydrostatic equilibrium at the intersections between the magnetic and rotational equators. Comparison with the available data (photometry and H$\alpha$ line profiles) showed  that the model provided a good quantitative description of the circumstellar environment.

%\begin{itemize}
%\item \textbf{Shall we show any model results here? We can show either Rich's or HDUST model results, showing the large discrepancy between model and observations. No attempt would be made to explain the differences, but we could discuss that the so far missing mechanism for venting the magnetosphere might be the solution for this.}
%\end {itemize}

In an attempt to reproduce the observed polarization modulation of $\sigma$\,Ori\,E, we fed the predicted density distribution of the RRM model of \citet{2005ApJ...630L..81T}  to the HDUST radiative transfer code \citep{car06}. Since all stellar and geometrical parameters were taken from \citet{2005ApJ...630L..81T}, the only free parameter in the model is the density scale of the magnetosphere. %, which was adjusted so that the model correctly reproduces the depth of the photometric light curve.

The results of this modeling are shown in Fig.~\ref{fig:rrm_fit}. We could not find a model that matched simultaneously both the photometry and the polarimetry.
A higher density model (solid line) that matches the depth of the photometric eclipses predicts a polarization amplitude that is three times larger than what is observed. Conversely, a lower density model that reproduces the amplitude of the polarization fails to reproduce the photometric amplitude. %Also, both models predict very large variations in the polarization position angle that are not observed. 

Furthermore, the detailed shape of the polarization curve, in particular the position angle variation, is not well matched. Since the  polarization postion angle is sensitive to the geometry of the scattering material, this discrepancy indicates that the primary difficulty with the basic RMM model is the shape of its density distribution.

\begin{figure}[!t]
\centering
\includegraphics[trim=105 120 90 80,clip=true,width=.45\linewidth]{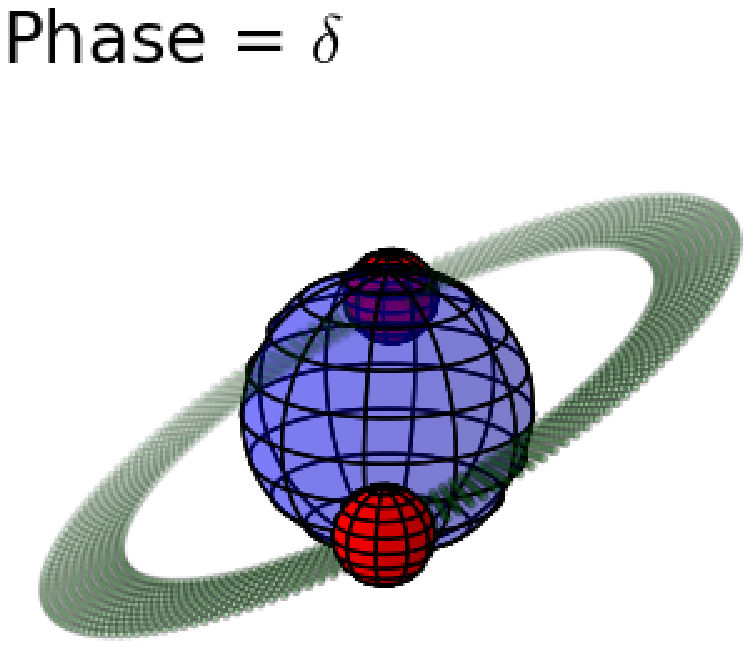}
\includegraphics[trim=105 120 90 80,clip=true,width=.45\linewidth]{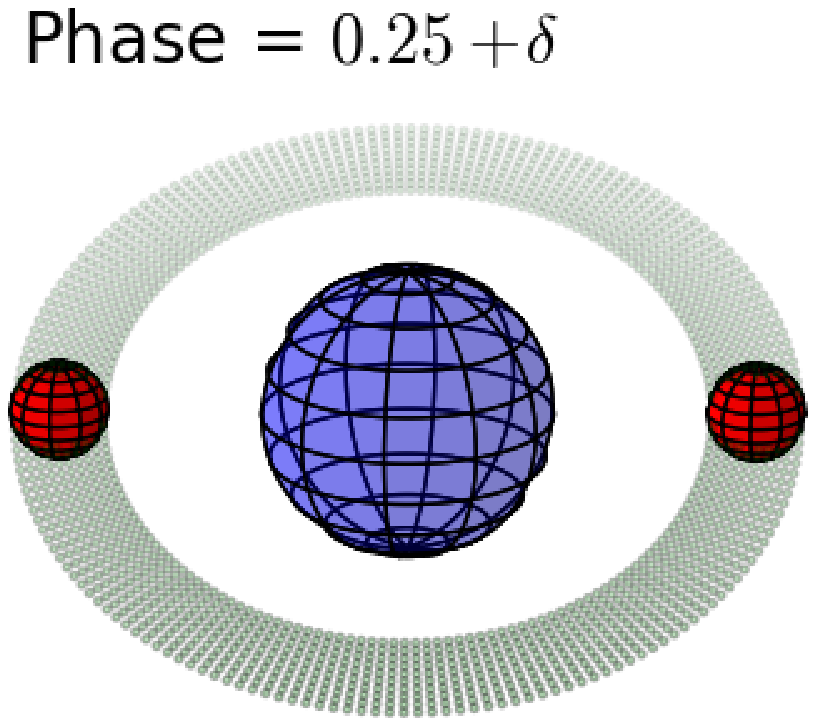}
\includegraphics[trim=105 120 90 80,clip=true,width=.45\linewidth]{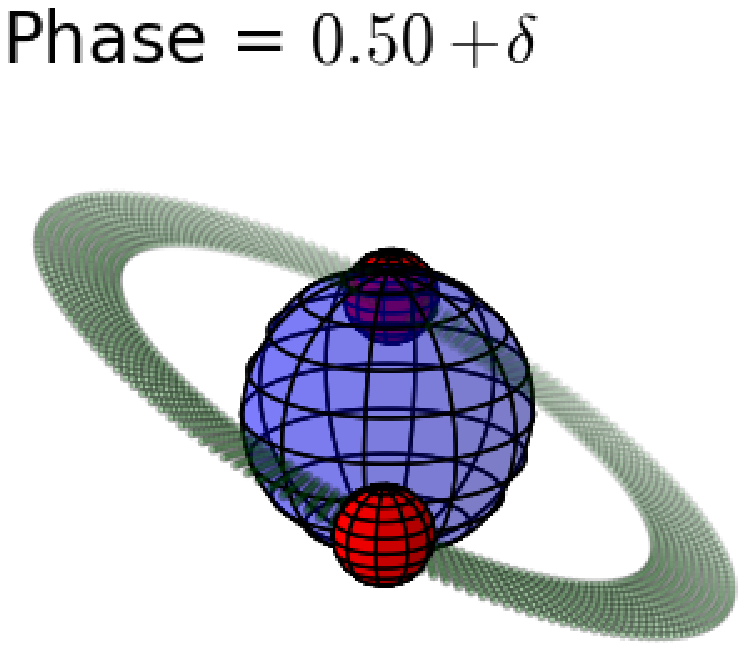}
\includegraphics[trim=105 120 90 80,clip=true,width=.45\linewidth]{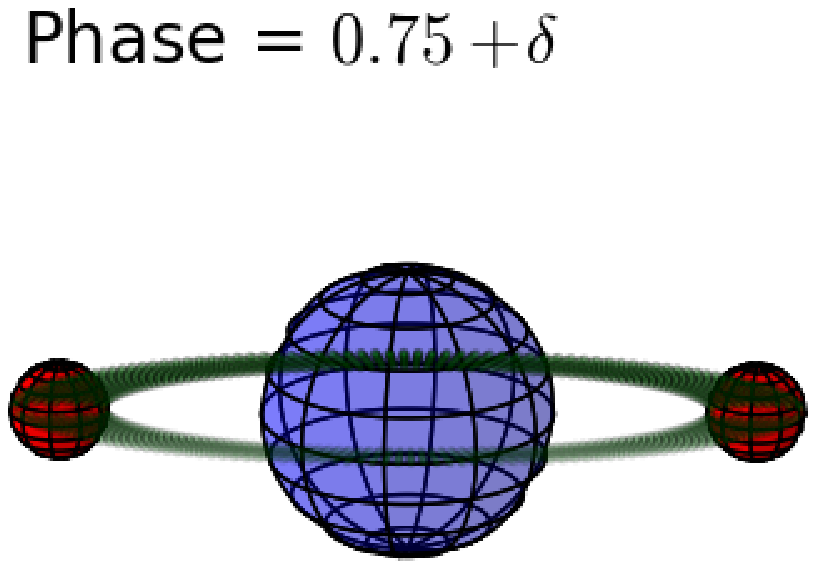}
\caption{Geometric conception of the ``dumbbell + disk'' model, to scale. \label{fig:model} }
\end{figure}

%%%%%%%%%%%%%%%%%%%%%%%%%%%%%%%%%%%%%%%%%%%%%
%%%%%%%%%%%%%%%%%%%%%%%%%%%%%%%%%%%%%%%%%%%%%
%%%%%%%%%%%%%%%%%%%%%%%%%%%%%%%%%%%%%%%%%%%%%
\section{Single Scattering Model} \label{sec:dumbbell}

We now explore what changes must be made to the basic RRM model to better reproduce the shape of the polarimetric variability. The
RRM predicts the existence of diametrally opposed plasma clouds confined in the rotational equator and a diffuse plasma disk close to the magnetic equator. Based on this idea, we developed a simple toy model for $\sigma$\,Ori\,E that consists of two elements: 1) a thin uniform disk tilted by an angle $\psi$ from the rotational equator, and 2) a pair of spherical blobs situated in the equatorial plane at the intersection between this plane and the magnetic equator, thus forming a ``dumbbell-like'' structure. This configuration is illustrated in Fig~\ref{fig:model}, for several  orbital phases and for a viewing angle of $70^\circ$.

Since the polarization is small, we may perform the calculation of the $Q$ and $U$ Stokes parameters as a function of orbital phase for a given inclination angle $i$ following a simple single-scattering approach (valid only in the optically thin limit). The scattered flux is first determined in the frame of the star using the formalism described in \citet[Eq. 6]{1994ApJ...436..818B} and then rotated to the frame of the observer \citep[Eq. 20]{1994ApJ...436..818B}.  

For this initial study, we adopted {5 fixed parameters}, % to describe the blobs' geometry. 
listed in Table~\ref{tab:mod}, whose values were based on the models of \citet{2005ApJ...630L..81T}. They are the stellar radius, $R_{\star}$, the distance of the blobs from the star, $d_{\rm b}$, the size of the blobs, $R_{\rm b}$, the geometrical thickness of the disk, $H_{\rm d}$, and the inclination angle, $i$.
The adopted value for $R_{\rm blob}$ is somewhat arbitrary, since in the limit of $R_{\rm blob} \ll R_{\star}$ and in the single-scattering approximation, a large, tenuous blob is roughly equivalent to a dense, small blob. 
For the same reason, the disk geometrical thickness is also somewhat arbitrary: if the single-scattering approximation holds, a thinner, denser disk is equivalent to a thicker, less dense disk. In other words, what really controls the polarization level of each component is their total scattering mass.
%\addDMF{For the disk, we kept the same position and size of the blobs.}
For the inclination angle we adopt two possible values: $i=70^{\circ}$, corresponding to a projected counterclockwise rotation of the blobs on the sky and $180^\circ-70^\circ=110^\circ$, corresponding to a clockwise rotation.

In this analysis we made \emph{no previous assumptions about the IS polarization}; i.e., we fitted directly the observed $Q$ and $U$ Stokes parameters. %shown in Fig.~\ref{fig:observed}. 
The adopted modeling procedure is as follows:
\begin{enumerate}
\item Choose values for electron number density in the blob and disk ($n_e^{\rm b}$ and $n_e^{\rm d}$), the phase lag, $\delta$, between the photometric primary minimum ($\phi=0$) and the orbital phase where the blobs are aligned with the line of sight, the tilt angle of the disk, $\psi$, and the position angle of the equatorial plane in the plane of the sky,  $\theta$. These 5 parameters define a unique model for the intrinsic $Q$ and $U$ in the equatorial frame.
\item Choose values for $Q_{\rm IS}$ and $U_{\rm IS}$ and use them to calculate synthetic observed $Q$ and $U$ parameters.
\item Calculate the reduced $\chi^2$ of the model.
\item Repeat steps 1 and 3 for several hundred of values for the {7 free parameters} and choose the model with minimum $\chi^2$.
\end{enumerate}

The parameters of our best-fitting model {(reduced $\chi^2=2.8$)} are listed in Table~\ref{tab:mod}. The estimated errors are for a 95\% confidence interval. A notable result is that the IS polarization, independently obtained by this analysis, is consistent (within the errors) with the estimate made using $\sigma$\,Ori\,AB as a measure of the IS. Both the orientation of the equatorial plane in the sky ($\theta$) and the phase lag ($\delta$) are consistent with the estimate made in Sect.~\ref{results_aca}. The resulting blob density corresponds to an electron optical depth of  {$\tau=0.13$}, %\addDMF{!Nao mudou!}, 
which does not violate the  optically thin assumption. It is worth mentioning that our estimate for the electron density in the blobs, $10^{12}\,e^-$\,cm$^{-3}$, is in good agreement with the values derived by \citet{1982A&A...116...64G} and \citet{2007A&A...475.1027S} using the Inglis-Teller formula. {From the best-fitting parameters one can calculate the mass of each model component. Assuming a molecular weight of 0.6, which corresponds to a fully ionized gas with solar chemical composition, the mass of each blob is $M_{\rm b} = 6.0\times 10^{-12}\,M_\odot$ and the mass of the disk is $M_{\rm d} = 1.0\times 10^{-11}\,M_\odot$.
Therefore, we obtain that both components have similar masses ($2M_{\rm b}/M_{\rm d}=1.2$).}

In Fig.~\ref{fig:intrinsic}, right plot,  we show the observations corrected by the IS polarization of Table~\ref{tab:mod}. The solid curves correspond to the best-fitting single scattering model. The two minima of the model polarization occur at the phases for which the two blobs are alined with the line of sight. %The model polarization is close to zero at these phases because of both the spherical geometry of the blob and the near edge-on viewing angle.
At these phases, the model $Q$ parameter is minimum and the $U$ parameter is maximum (negative for $\phi \approx \delta$ and positive for $\phi \approx 0.5+\delta$), because the disk dominates the polarized flux at these phases. Conversely, the polarized flux is dominated by the blobs at the phases at which the blobs are on the side of the star ($\phi \approx 0.25+\delta$ and positive for $\phi \approx 0.75+\delta$).

An interesting result is that the data allowed us to firmly establish that the blobs are rotating counterclockwise on the plane of the sky.  
In the right plot of  Fig.~\ref{fig:intrinsic}, the solid curves show the best-fitting model for $i=70^\circ$ (reduced $\chi^2=2.8$) while the dotted curves correspond to the best fitting model with $i=110^\circ$ (reduced $\chi^2=5.1$). The former is clearly a much better fit to the data.

%{The large value of the reduced $\chi^2$ indicates that the single scattering blob model is  insufficient to reproduce the polarimetric data (mainly in the second half of the orbital period), but the general agreement between model and data (Fig.~\ref{fig:intrinsic}), together with the fact that $Q_{\rm IS}$, $U_{\rm IS}$, $\theta$ and $\delta$ are consistent with the independent determinations made on Sect.~\ref{results_aca}, is indicative that this simple ``dumbbell'' model can be a useful tool for an initial analysis of the polarization data of other magnetic stars.}

\begin{figure}[!t]
\centering
\includegraphics[width=.9\linewidth]{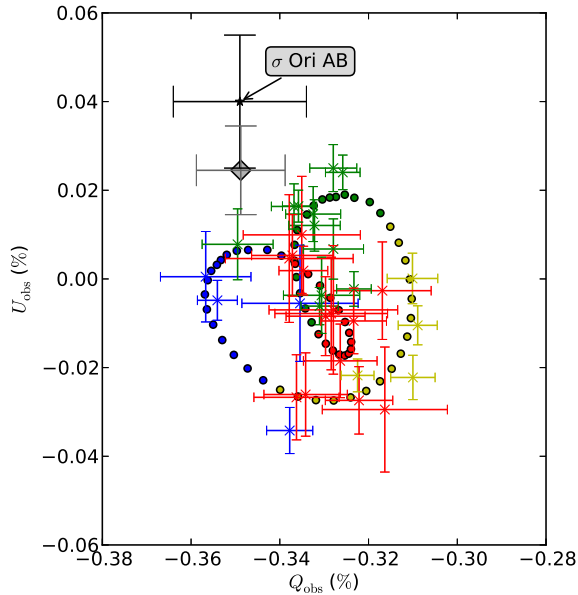}
\caption{$QU$ plot of the observed data (points with error bars). The orbital phase correspondence of the points is as follows: blue, $\delta$--$(\delta+0.25)$, orange, $(\delta+0.25)$--$(\delta+0.50)$, green, $(\delta+0.50)$--$(\delta+0.75)$, and red, $(\delta+0.75)$--$\delta$.
The filled circles show the best-fitting model of Table~\ref{tab:mod}. The IS polarization of $\sigma$\,Ori\,AB is indicated. The grey diamond shows the IS of Table~\ref{tab:mod}.
\label{fig:qu} }
\end{figure}

{An alternate comparison between the data and the single-scattering model is made in Fig.~\ref{fig:qu}, where we plot the observed (i.e., uncorrected for IS polarization)  $U$ vs. $Q$ parameters. %The data was binned to increase the. 
%To facilitate the visualization of the phase correspondence of the data, 
The blue points correspond to the phase range $\delta$--$(\delta+0.25)$, the orange ones from $(\delta+0.25)$--$(\delta+0.50)$, the green ones from  $(\delta+0.50)$--$(\delta+0.75)$ and the red ones from $(\delta+0.75)$--$\delta$.
In the $QU$ plane the behavior of the data is quite different from the first half-cycle to the second: while in the first half  (blue and orange points) the data forms a  nearly-circular outer loop, in the second half (green and red points) there is an inward incursion.
The model reproduces this behavior.

It is important to point out that the two-component model (blob+disk) adopted by us is required to produce such a behavior in the $QU$ plane. For instance, a model with only the blobs would form a symmetric double ellipse in the $QU$ plane, while a model with only the tilted disk would result in a shape resembling a crescent moon. The observed curve, composed of an outer loop and an inner incursion, can only be obtained by a combination of these two components.}

\begin{deluxetable}{llc}

%% This is the title of the table.
\tablecaption{Parameters of the single scattering model}

%% The \tablehead gives provides the column headers.  It
%% is currently set up so that the column labels are on the
%% top line and the units surrounded by ()s are in the 
%% bottom line.  You may add more header information by writing
%% another line between these lines. For each column that requries
%% extra information be sure to include a \colhead{text} command
%% and remember to end any extra lines with \\ and include the 
%% correct number of &s.
\tablehead{\colhead{Parameter} & \colhead{Value} & \colhead{Type} \\ 
\colhead{} & \colhead{} & \colhead{} } 

%% All data must appear between the \startdata and \enddata commands
\startdata
$i$ (degree) & $70$ or $110$ & fixed \\
$R_{\star}$ ($R_\odot$) & $4.3$ & fixed \\
$d_{\rm b}$ ($R_{\star}$) & $2.4$ & fixed \\
$R_{\rm b}$ ($R_{\star}$) & $1/3$ & fixed \\
$H_{\rm d}$ ($R_{\star}$) & $0.01$ & fixed \\
$n_e^{\rm b}$ ($e^-$\,cm$^{-3}$) & $1.0^{+0.6}_{-0.9} \times 10^{12}$ & fitted \\
$n_e^{\rm d}$ ($e^-$\,cm$^{-3}$) & $2.7\pm 1 \times 10^{12}$ & fitted \\
$\psi$ (degree) & $28$ & fitted \\
$Q_{\rm IS}$(\%) & $-0.35 \pm 0.01$ & fitted  \\
$U_{\rm IS}$ (\%) & $0.025 \pm 0.010$ & fitted \\
$\theta$ (degree) & $150\pm7$ & fitted \\
$\delta$ & $-0.17\pm0.02$ & fitted 
\enddata

%'% Include any \tablenotetext{key}{text}, \tablerefs{ref list},
%% or \tablecomments{text} between the \enddata and 
%% \end{deluxetable} commands

%% No \tablecomments indicated

%% No \tablerefs indicated

\label{tab:mod}

\end{deluxetable}

\section{Conclusions \label{sec:discussion_aca}}

The polarimetric data presented here for $\sigma$\,Ori\,E has an accuracy of at least one order magnitude better than those shown in the work of \citet{1977ApJ...218..770K} and thus represent the first firm detection of the polarization modulation of a co-rotating magnetosphere. The intrinsic polarization, calculated using $\sigma$\,Ori\,AB as a measure of the IS polarization, displays a roughly symmetric sinusoidal curve ranging from 0.02 to 0.08\%, with a period of half of the orbital period and a phase lag of -0.085 between the polarization minimum and the primary minimum of the light curve. {The polarization position angle is also variable. Its mean value of $150\fdg0$ indicates that the minor elongation axis of the asymmetric structure around $\sigma$\,Ori\,E is oriented $150\fdg0$ east of the celestial north.}

A preliminary analysis made with the RRM model \citep{2005ApJ...630L..81T} suggests that it is difficult to fit both the photometry and the polarization simultaneously. This seems to imply that the geometry of the plasma clouds, as predicted by the current version of the model, is incorrect. % High precision polarimetry, which provides a diagnostic of the geometry of clouds,  serves as useful constraints for the RRM model. 

{However, the data was well reproduced by a  two-component model, consisting of two spherical, diametrally opposed blobs and a disk tilted with respect to the rotational equator. 
{
%This model was physically motivated by the RRM model that predicts higher densities at the minimum of the effective potential, which is at the magnetic equator, with highest densities where the magnetic equator intersects the rotational equator; i.e., an inclined disk with density enhancements at the line of nodes -- the blobs.
This model was physically motivated by the RRM model, which predicts accumulation of centrifugally supported wind plasma in a disk-like structure with the highest densities (corresponding to the blobs) at the points where the magnetic equator intersects the rotational equator.}
This model provided useful constraints on the total scattering mass of each component. Also,  it provided an independent estimate for the IS polarization, as well as for the orientation of the equatorial direction in the plane of the sky. 
{Our best-fitting model suggests that the mass of each blob is $M_{\rm b} = 0.6\times 10^{-12}\,M_\odot$ and the mass of the disk is $M_{\rm d} = 1.0\times 10^{-11}\,M_\odot$, which indicates that both components have similar masses ($2M_{\rm b}/M_{\rm d}=1.2$).}
Also, the tilt angle of the disk with respect to the plane of the sky is $28\deg$, similar to the angle between the rotational and magnetic axes  \citep{2005ApJ...630L..81T}.} {Finally, the model allowed us to determine  that the blobs are rotating counterclockwise on the plane of the sky.}

%The analysis with the ``dumbbell'' model also allowed us to establish that the rotation direction of the blobs in the plane of the sky is clockwise. 
%The good agreement between model and data (Fig.~\ref{fig:intrinsic}), together with the fact that $Q_{\rm IS}$, $U_{\rm IS}$, $\theta$ and $\delta$ are consistent with the independent determinations made on Sect.~\ref{results_aca}, is indicative that this simple ``dumbbell'' model can be a useful tool for an initial analysis of the polarization data of other magnetic stars.

%However, the model fit is not satisfactory, mainly in the second half of the orbital period ($\phi>0.5$), which indicates that a spherical geometry is probably a poor description for the clouds' geometry.  Despite being unable to reproduce the details of the observations, 

\acknowledgments
A.C.C. acknowledges support from CNPq (grant 308985/2009-5) and Fapesp (grant 2010/19029-0).
D.M.F. acknowledges support from FAPESP (grant 2010/19060-5).
R.H.D.T. acknowledges support from NSF grants AST-0908688 and AST-0904607, and NASA grant NNX12AC72G.
This work has made use of the computing facilities of the Laboratory of Astroinformatics (IAG/USP, NAT/Unicsul), whose purchase was made possible by the Brazilian agency FAPESP (grant 2009/54006-4) and the INCT-A.

\end{document}